# Rapid, antibiotic incubation-free determination of tuberculosis drug resistance using machine learning and Raman spectroscopy

**Authors:** Babatunde Ogunlade[1]*†, Loza F. Tadesse[2,3,4]*, Hongquan Li[5]*, Nhat Vu[6], Niaz Banaei[7], Amy K. Barczak[4,8,9], Amr. A. E. Saleh[1,10], Manu Prakash[2] and Jennifer A. Dionne[1,11]†

**Affiliations:**
[1] Department of Materials Science and Engineering, Stanford University; Stanford, 94305, CA, USA.

[2] Department of Bioengineering, Stanford University School of Medicine and School of Engineering; Stanford, 94305, CA, USA.

[3] Department of Mechanical Engineering, Massachusetts Institute of Technology; Cambridge, 02142, MA, USA.

[4] The Ragon Institute, Massachusetts General Hospital; Cambridge, 02139, MA, USA.

[5] Department of Applied Physics, Stanford University; Stanford, 94305, CA, USA.

[6] Pumpkinseed Technologies, Inc; Palo Alto, 94306, CA, USA.

[7] Department of Pathology, Stanford University School of Medicine; Stanford, 94305, CA, USA.

[8] Division of Infectious Diseases, Massachusetts General Hospital; Boston, 02114, MA, USA.

[9] Department of Medicine, Harvard Medical School; Boston, 02115, MA, USA.

[10] Department of Engineering Mathematics and Physics, Cairo University; Giza, 12613, Egypt.

[11] Department of Radiology, Molecular Imaging Program at Stanford (MIPS), Stanford University School of Medicine; Stanford, 94035, CA, USA.

*Indicates equal contribution

† To whom correspondence should be addressed;

E-mail: bogun@stanford.edu; jdionne@stanford.edu.

**Author contributions:**

- Conceptualization: LFT, AAES, BO, HL, AKB, MP, JAD
- Methodology: LFT, BO, HL, AAES, NV
- Investigation: BO, HL, LFT
- Formal Analysis: BO, NV, HL
- Visualization: BO

- Funding acquisition: JAD, AKB, MP
- Resources: JAD, AKB, MP, NB
- Writing – original draft: LFT, BO, HL, AAES, JAD
- Writing – review & editing: LFT, BO, HL, AKB, NV, JAD

**Competing interests:**

J.D. and N.V. are shareholders in Pumpkinseed Technologies.

**This PDF file includes:**
**Main Text**
**Figures 1 to 4**

**Abstract:**
Tuberculosis (TB) is the world's deadliest infectious disease, with over 1.5 million deaths annually and 10 million new cases reported each year[1]. The causative organism, *Mycobacterium tuberculosis* (Mtb) can take nearly 40 days to culture[2,3], a required step to determine the pathogen's antibiotic susceptibility. Both rapid identification of Mtb and rapid antibiotic susceptibility testing (AST) are essential for effective patient treatment and combating antimicrobial resistance. Here, we demonstrate a rapid, culture-free, and antibiotic incubation-free drug susceptibility test for TB using Raman spectroscopy and machine learning. We collect few-to-single-cell Raman spectra from over 25,000 cells of the MtB complex strain Bacillus Calmette-Guérin (BCG) resistant to one of the four mainstay anti-TB drugs, isoniazid, rifampicin, moxifloxacin and amikacin, as well as a pan-susceptible wildtype strain. By training a neural network on this data, we classify the antibiotic resistance profile of each strain, both on dried samples and in patient sputum samples. On dried samples, we achieve >98% resistant versus susceptible classification accuracy across all 5 BCG strains. In patient sputum samples, we achieve ~79% average classification accuracy. We develop a feature recognition algorithm in order to verify that our machine learning model is using biologically relevant spectral features to assess the resistance profiles of our mycobacterial strains. Finally, we demonstrate how this approach can be deployed in resource-limited settings by developing a low-cost, portable Raman microscope that costs <$5000. We show how this instrument and our machine learning model enables combined microscopy and spectroscopy for accurate few-to-single-cell drug susceptibility testing of BCG.

**Significance statement:**
Tuberculosis is preventable and curable, yet it is the world's deadliest infectious disease. This is in part due to the emergence of drug-resistant tuberculosis. Timely, accurate drug susceptibility testing of *Mycobacterium tuberculosis* is critical for effective patient treatment and the prevention of community spread, as inappropriate usage of anti-TB drugs may delay treatment progress and generate acquired resistance. Here, we develop a fast antibiotic susceptibility test for tuberculosis by combining Raman spectroscopy and machine learning. Using this methodology, we can assess the resistance of antibiotic-resistant mycobacteria strains at clinically relevant speeds and accuracies. Our findings provide a foundation for rapid, low-cost, point-of-care mycobacterial drug susceptibility testing for diagnostic and surveillance applications, a major step in the fight against antimicrobial resistance.

**Main text:**

## INTRODUCTION

The discovery of antibiotics in the early 20th century marked a turning point in our defense against tuberculosis (TB) – one of the deadliest infectious diseases known to humans. At that time, TB had become curable and was considered to be on a path toward elimination. However, by the end of the 20th century, TB re-emerged as a leading cause of death globally, in part due to challenges in diagnosis and its evolved resistance to antibiotics. While growth in liquid culture is considered the gold standard for organism identification and determination of antibiotic susceptibility, the causative organism, *Mycobacterium tuberculosis* (Mtb) can take up to 40 days to culture. As antibiotic susceptibility testing requires pathogen growth, definitive determination of Mtb antibiotic resistance using traditional antibiotic susceptibility testing (AST) can take additional weeks. In the delay between diagnosis and AST results, many patients are treated with only partially active antibiotic regimens, giving rise to resistant strains. In 2021 alone, more than half a million cases of multidrug resistant TB infections were reported[1].

To mitigate the rise of antimicrobial resistance, a key strategy of the World Health Organization (WHO) is to improve surveillance of antibiotic-resistant infections, and to promote the appropriate use of quality medicines[2]. To address WHO guidelines, several culture-free AST approaches are being developed. For example, nucleic acid amplification tests and polymerase chain reaction (PCR) do not require culturing and are highly sensitive and specific. However, these tests are limited to identifying resistance in cases in which a defined number of known genomic mutations confer resistance (e.g., the rpoB gene for resistance to rifampicin[3]). In general, the appearance of genetic mutations in previously identified genes of interest have not been consistently linked to resistance to these antibiotics. Moreover, known genetic elements or mutations typically constitute only a fraction of the genetic basis for clinically relevant resistance[4]. These assays also require multiple reagents, can suffer from errors with each thermal cycle, and cannot distinguish between live and dead bacteria, so they cannot be used to monitor treatment efficacy. Further, these assays cost well above the targeted $5000 in capital cost specified by the WHO for next generation TB drug-susceptibility testing tools[5]. Complementing PCR, loop-mediated isothermal amplification (LAMP) assays eliminate the need for thermal cycling. However, these LAMP assays cannot detect mutations in resistance-associated genes because of their inability to resolve single-nucleotide differences[6]. Rapid lateral flow point-of-care test antigen/antibody detection methods provide simpler and faster alternatives to LAMP and PCR but have poor sensitivity and specificity. There is also a significant delay in the appearance of target antigens in sputum or antibodies in the bloodstream, and these antigens do not necessarily indicate Mtb drug susceptibility. Therefore, current AST methods do not collectively provide the speed, sensitivity, and specificity needed in one system to meet WHO's goals for drug-resistant TB eradication.

Raman spectroscopy has the potential to identify the antibiotic resistance profiles of bacteria at the few-to-single-cell level using acquisition times on the order of seconds. Raman spectroscopy utilizes inelastic light scattering to probe the vibrational modes of a sample as a fingerprint. Different bacterial phenotypes are characterized by unique biomolecular compositions, leading to subtle differences in their corresponding Raman spectra. The Raman spectra of biological macromolecules typically reside between 500 and 1900 $cm^{-1}$, with lipids, proteins, and nucleic acids exhibiting fingerprints between 1000-1700 $cm^{-1}$, 1200-1660 $cm^{-1}$, and 600-1100 $cm^{-1}$, respectively[7]. As such, these unique Raman spectral signatures can be used for accurate cellular detection, identification, and antibiotic susceptibility testing. Using Raman spectroscopy and machine learning (ML) based spectral analysis, we have previously shown that genetically engineered antibiotic resistant and susceptible *S. aureus* strains can be classified with ~89% accuracy[7]. Similarly, we and others have used machine learning-assisted Raman to identify over 31 bacterial species and strains[8,9,10,11], including pathogens in liquid solvents[12] and in and blood[13,14,15]; to identify substrains of *E.coli*[16,17], to identify various respiratory viruses[18], to identify various cellular metabolites and secretomes[19], to assess the effects of antibiotics on resistant and susceptible bacteria[20,21], and to determine *S. aureus* antibiotic susceptibility without antibiotic incubation[22]. Despite these pioneering studies, the robust determination of antibiotic susceptibility of TB, and methods translation to TB-endemic regions, remains an open challenge. Previous studies have demonstrated the use of Raman and machine learning for mycobacterial detection and antibiotic resistance profiling with high classification accuracies[23,24]. However, the machine learning models trained in these studies behave like black boxes, lacking transparency in their decision-making and interpretability and raising questions concerning the spectral features being used to achieve such high classification accuracies. These questions are especially salient when spectra are collected in complex physiological media such as sputum, which has significant spectral overlap with the vibrational information contained in TB.

Here, we demonstrate a rapid, culture-free, and antibiotic incubation-free antibiotic susceptibility test for TB, based on the integration of few-to-single-cell Raman spectroscopy and machine learning. In particular, we systematically classify the antibiotic resistance profile of 5 isogenic Bacillus Calmette-Guérin (BCG) strains using a convolutional neural network. We collect Raman spectra from over 25,000 individual cells which are controllably engineered to possess resistance to one of the four mainstay anti-TB drugs (first line drugs isoniazid and rifampicin and second line drugs moxifloxacin and amikacin) and a pan-susceptible wild type strain. Using only 15-second integration times and Raman features from across the entire spectral range (740-1802 $cm^{-1}$), we achieve >98% resistant versus susceptible classification accuracy across all 5 BCG strains. In addition, we show that various genetic mutants can be accurately classified according to their antibiotic resistance. Using our feature recognition algorithm, we also identify the subsets of wavenumbers which most strongly influence antibiotic resistance classification and correlate them with the Raman signatures of well-known biomolecules and functional groups, verifying that our machine learning model is using biologically relevant peaks for classification. Our algorithm shows that antibiotic resistance behavior is primarily reflected in certain vibrational modes – particularly those of mycolic acid - that underly the molecular-level antibiotic response of the bacteria. This algorithm also shows that targeting the entire spectral range of scientific grade Raman instruments is not necessary for TB AST. To show how our approach can be robustly deployed in resource-limited settings, we develop a low-cost portable Raman setup for point-of-care applications. Our instrument costs <$5000, and with inclusion of low-cost plasmonic nanoantennas for surface-enhanced Raman spectroscopy (SERS), achieves an average accuracy of 89% across all 5 BCG strains. In dried patient sputum samples, our instrument achieves ~79% classification accuracy – providing a foundation for rapid, low-cost, and portable AST testing in resource-limited regions.

## RESULTS

### Few-to-single-cell Raman spectroscopy of BCG for antibiotic incubation-free determination of drug resistance

As summarized in Figure. 1, we deposit mycobacterial cells onto gold-coated glass substrates and collect few-to-single-cell Raman spectra (Fig. 1A). For our studies, we controllably engineer antibiotic-resistant Mtb models using BCG, a live-attenuated form of *Mycobacterium Bovis* as a model organism. While lacking central virulence determinants, BCG is nearly identical to Mtb in its core bacterial functions, including DNA replication, RNA polymerase, and cell wall functions We use these models to controllably engineer single-antibiotic resistance for robust profiling and to enable safe handling of resistant samples in our BSL-2 facilities for the proof-of-concept studies performed here. Individual Raman spectra are used as an input into our convolutional neural network; we employed a 10-fold cross validation scheme to determine the unbiased model performance. The output of this algorithm is a series of probability scores for antibiotic resistance (Fig. 1B). The class with the highest probability score is chosen as the predicted class.

We performed our antibiotic susceptibility test on five isogenic BCG strains: those resistant to the first-line TB antibiotics isoniazid and rifampicin; those resistant to the second-line TB antibiotics moxifloxacin and amikacin; and also on a fully antibiotic-susceptible control strain (hereby described as wildtype). Figure 1C shows the minimum inhibitory concentration of the four anti-TB antibiotics used on our five BCG strains, confirming each resistant strain's resistance to a single antibiotic as well as the full antibiotic susceptibility of the wildtype strain. As seen in Figures 1D-H, scanning electron micrographs show no distinct morphological difference among the resistant strains.

As shown in Fig. 2A, the average Raman spectra of the 5 distinct BCG strains (~ 1700 spectra per strain) look similar, with major spectral peaks preserved across the strains. However, there are some noticeable differences in the ~1000, ~1300, ~1500 and ~1700 $cm^{-1}$ bands between wildtype and four resistant strains; the majority of these vibrational modes correspond to cell-wall components such as mycolic acid and proteins. Therefore, the differences between the spectra are most likely due to differences in cell-wall composition and structure, discussed later in the text. We performed t-distributed stochastic neighbor embedding (t-SNE) which shows significant clustering for each of the 5 strains (Fig. 2B). Such clustering suggests that accurate classification by resistance should be possible. To classify the antibiotic resistance profiles of the 5 BCG strains, we developed a deep learning model, a ResNet[25] variant[8]. We perform a stratified K-fold cross validation of our convolutional neural network's performance across 10 splits. As seen in Figure 2C, we achieve an average accuracy of 98% across all BCG strains. Note that main diagonal elements represent correct classifications and off-diagonal elements represent misclassifications. High classification accuracies were also obtained with different gold-coated glass substrates (Fig. S3) as well as with combined datasets from different experimental replicates (Figs. S11 & S12). These results demonstrate that our approach is agnostic to the particular substrate selected or batch effects and are solely from intrinsic differences in the strains themselves. Therefore, Raman spectroscopy in conjunction with machine learning can be used to predict antibiotic resistance in mycobacteria without the need for antibiotic incubation and hence without culturing.

**Select spectral bands are most crucial for predicting antibiotic resistance**

We quantitatively identify the regions of the spectra that are key for the high classification accuracy. To identify these meaningful bands, we used a probing algorithm that perturbs the input data before re-running the classification through the trained models from Fig. 2C[26]. Using each test fold from our 10-fold cross validation, we iterate through the wavenumbers and at each iteration, perturb the spectrum by modulating the amplitude of the spectral intensity with a Voigt distribution centered at the probing wavenumber. After each perturbation, we recalculate the classification accuracy, compare the updated results with our baseline classification accuracy, and determine the importance for each wavenumber - the greater the decrease in accuracy due to a given perturbation, the more important the wavenumber. As shown in Fig. 2D, these regions are spectral bands located around 1295, 1437 and 1660 $cm^{-1}$ [27,28]. These wavenumber bands correspond to mycolic acid $-CH_2$ twist and $-CH_2$ deformation modes, and Amide I C=O stretching modes, which change when the bacteria develop resistance. Therefore, our analysis suggests that Raman spectral features can be used to differentiate resistant bacteria that exhibit distinct biochemical profiles.

Next, we reduced the wavenumber feature input to our neural network, reflecting those bands most important for high-accuracy classification. Using only the top 50 wavenumbers, we achieve ~86% accuracy. Classification accuracy increases to 96% when using 250 features, and further to 98% when using all 1930 features (Fig. 2E). This result indicates potential for more rapid spectral collection and analysis as full spectroscopic analysis is not necessary to achieve high classification accuracies.

**Antibiotic-resistant mutants can be accurately classified**

To test the robustness of our approach to potential patient-patient variation, we interrogated four distinct isoniazid-resistant mutants and three distinct amikacin-resistant mutants (Fig. S13). As seen in Figure 3A, the isoniazid-resistant mutants differ from each other by single point mutations in the katG gene, which encodes for catalase peroxidase, an enzyme that converts isoniazid to its biologically active form[29]. Isoniazid resistant mutant 1 has a single point mutation at position 609 from cytosine to thymine; isoniazid resistant mutant 2 has two single point mutations at positions 290 and 609 from adenine to guanine and cytosine to thymine, respectively; isoniazid resistant mutant 3 has a single point mutation at position 1292 from guanine to adenine; and isoniazid resistant mutant 4 has no single point mutations identified in katG, and likely harbors a non-katG mutation conferring the observed isoniazid resistance. As seen in Figure 3B, Raman spectra from these distinct isoniazid-resistant mutants are seemingly identical, with tightly overlapping projections on t-SNE clustering (Fig. 3C).

The three amikacin-resistant mutants reveal similar clustering on the t-SNE projection. Importantly, however, the clustering is distinct from the isoniazid cluster. The classification accuracy to predict the antibiotic susceptibility of each strain within its antibiotic class is ~99% (Fig. S2). Clinically, this result is significant, as it is likely that antibiotic-resistant mutants from different patients may differ genetically, and we are able to identify resistance to a particular antibiotic regardless of this difference.

**Enabling low-cost and portable Raman for Mtb drug testing**

To demonstrate that our findings can translate to point-of-care TB drug testing in resource-limited regions, we developed a low-cost, portable, and fully automated Raman microscope (Fig. 4A). This Raman microscope is based on the Octopi/Squid modular microscope framework developed by our team for fluorescence[30,31], but advances this setup for spectroscopy. The instrument utilizes a non-cooled CMOS camera (Sony IMX290), a transmission diffraction grating with >95% efficiency, and a design without a pinhole or slit for low-cost spectroscopy. A fiber-coupled 785 nm VHG-stabilized laser is used as the excitation source. To boost the signal from bacteria, we synthesize gold nanoparticles (AuNPs) for surface-enhanced Raman scattering. The Dionne Group has done substantial research using gold nanorods to enhance Raman signatures from bacterial and mammalian cells for both dried and liquid samples, showing that a large range of plasmon resonances from 670 to 860 nm can be used to obtain high quality SERS spectra for 785 nm excitation[12,13].

As seen in Figure 4B, the nanorods used in this study have a scattering resonance near the pump-laser wavelength. Compared to whole-cell Raman, these nanorods yield a ~2 order of magnitude increase in Raman scattering[10,12,13,32,33,34,35,36] from single cells and pathogens and considerably shorten the integration times required for few-to-single-cell analysis on our microscope. The AuNPs are also relatively inexpensive to synthesize or purchase, easy to tune optically[37,38,39] and easy to integrate with our bacterial strains. All components combined for this microscope and spectrometer are <$5000. Additional details on the sample preparation and imaging/spectroscopy are provided in SI.

As a baseline measurement, we first collected ~3000 spectra from the 5 BCG strains mixed with gold nanorods using 0.3 second integration times and 3 accumulations per sample (Fig. 4B). Here, the shortened integration time was chosen to maintain a similar signal-to-noise ratio of the signal as with the Raman of the scientific-grade instrument. Similar to the spectra collected on the scientific grade Raman microscope in Fig. 2A, the average Raman spectra of the 5 distinct BCG strains look similar, with major spectral peaks preserved across the strains. However, there are some noticeable differences in the ~1000, ~1300 and ~1400 $cm^{-1}$ bands between wildtype and four resistant strains (Fig. 4D). We performed t-SNE which shows significant clustering for each of the 5 strains (Fig. 4E). As seen in Figure 4f, we achieve an average accuracy of 89% across all 5 BCG strains, comparable with the ~95-98% average classification accuracy achieved on our scientific-grade tool.

As an initial evaluation of the performance for clinical samples, we spiked BCG pathogens into sputum (see Methods), mixed the sample with gold nanorods, and collected a total of ~5300 spectra from the BCG strains. Using only 0.3 second integration times and 3 accumulations per sample, we obtain an average classification accuracy of ~79% for the 5 strains across two separate sputum samples (Fig. 4I and Fig. S7). This accuracy is reasonable with that observed for bacteria alone on scientific grade Raman instruments (Fig. 2C) while providing rapid results at a lower cost and smaller footprint.

We note that the spectra differ from those without using AuNPs, which is as expected as now the Raman signatures primarily originate from the surface-bound molecules on the bacteria, due to the electric field localization of the AuNPs. This intraclass spectral variation is evident in the milder, less intense clustering of the 5 strains in the t-SNE in Figure 4E and Figure 4H compared to the whole-cell Raman collected in Figure 2B. However, our convolution neural network can still classify strains, as demonstrated by the high accuracies we are able to obtain. Future work spanning more patient data and optimized spectral data processing could enable clinical BCG strain classification.

## DISCUSSION

In summary, we demonstrate a rapid and accurate TB AST methodology that is antibiotic incubation-free and culture-free. Unlike current culture-based and nucleic acid amplification-based testing (NAATs) platforms, our Raman approach does not require lengthy culturing, expensive reagents, or thermal cycling equipment, and is robust to single nucleotide point mutations in resistance genes of interest. Our approach can accurately classify resistance across a variety of antibiotics, and across a variety of genetic mutations. In addition, our machine learning models not only provide accurate AST classification, but also interpretability to the Raman spectra. Indeed, as highlighted by our feature-recognition algorithm, resistance behavior is primarily reflected in certain vibrational modes, particularly those of mycolic acid, that underly the molecular-level antibiotic response of the bacteria. We developed a low-cost portable Raman microscope, capable of high BCG classification accuracy on dried samples and in dried sputum. To extend these results beyond that of the BCG laboratory strain, on-going work is aimed at testing our approach in the field on Mtb-positive patients. Specifically, we are planning to fold Octopi-Raman deployment into the growing Octopi infrastructure, which currently includes deployment all across the globe, including India, Nigeria, Senegal, Liberia, Kenya, and the United Kingdom. Such clinical work will elucidate the intrinsic differences between strain types, well beyond sputum-to-sputum variations. If successful, this clinical work will provide a new and needed approach to enable the TB surveillance and diagnostic milestones set by the WHO. Finally, our study does have limitations: Our work is proof of concept that probes the phenotypic differences of singly antibiotic-resistant laboratory BCG strains using Raman spectroscopy. However, further work should be done using this methodology to probe multidrug resistant BCG and Mtb. With more work in this direction, our methodology could be applied to curbing the growing global crisis of multidrug-resistant TB.

**Materials and Methods:**

BACTERIAL STRAINS

To raise antibiotic resistant mutants, BCG was plated on 7H10 media with increasing concentrations of each antibiotic of interest. Plates were incubated at 37 °C for 21 days. Individual colonies were selected and grown in 7H9 media containing OADC, 0.2% glycerol, and 0.05% Tween-80; antibiotic resistance was confirmed by Minimum inhibitory concentration (MIC) testing.

MIC DETERMINATION

Each strain was grown to mid-log phase in 7H9 media containing OADC, 0.2% glycerol, and 0.05% Tween-80. Cultures were then diluted back to an OD600 of 0.01 in media containing 2-fold dilutions of each antibiotic of interest. Cultures were grown at 37°C in 96-well plates; bacterial cells were mixed at 7 and 14 days of incubation, and growth was determined by OD600 measurements on a Tecan Spark plate reader.

BACTERIAL PREPARATION FOR SPECTROSCOPY

To obtain the antibiotic-resistant BCG strains, wildtype BCG (ATCC 35737 TMC 1019 [BCG Japanese]) was grown to mid-log phase in 7H9 liquid media supplemented with OADC, glycerol, and tween-80, then pelleted and plated on 7H10 solid media containing serial dilutions of each antibiotic of interest from 10-fold above the MIC to 10-fold below the MIC. Plates were incubated at 37°C for 4 weeks; individual colonies were selected for culture from the serial dilutions that yielded between 1 and 25 colonies. One wildtype colony was selected from the wildtype plate, one rifampicin-resistant colony was selected from the rifampicin plate, four amikacin-resistant colonies were selected from the amikacin plate, five isoniazid-resistant colonies were selected from the isoniazid plate, and five moxifloxacin-resistant colonies were selected from the moxifloxacin plate. Each colony had its antibiotic resistance profile confirmed using MIC and each colony was confirmed to be unique using whole genome sequencing and MIC, and hence were used as sub-strains (Supplementary Figure 13).

To determine MIC, wild-type BCG and each resistant strain was grown to mid-log phase, then diluted back to an OD600 of 0.01 and incubated in two-fold serial dilutions of each antibiotic in a 96-well plate. Days 7 and 14 post-inoculation, each well was gently mixed and OD600 was read on a TECAN Spark device. Only confirmed resistant mutants, as well as the fully susceptible (wildtype) strains were used for Raman testing. 1 mL of frozen stock of the wildtype and the confirmed resistant mutants were seeded in 10 mL 7H9 liquid media supplemented with OADC, glycerol, and tween-80 and incubated at 37 $^{\circ}$C shaking at 115 RPM until mid-log phase (~5-6 days) using the Thermo Scientific MaxQ 4450 incubator. Cultures were then chemically inactivated overnight in a 1:1 solution using 10% formalin. 1.5 mL of culture (~4 $*10^{8}$ cells/mL) was washed with water three times at 6000 rpm for 4 min using a mySPINTM 6 Mini Centrifuge. The BCG strains were prepared for Raman interrogation by resuspending the final washed pellet in 100 μL of sterile water and drying 3 μL of the suspension on a gold-coated glass substrate. Droplets were dried for 20 minutes, and Raman was collected immediately after drying.

For the sputum spiked BCG Raman interrogation, 25 μL of 3x washed bacteria was spun down and resuspended in 50 μL of sputum. After sitting for 1 minute, the spiked sputum was then spun down, washed once with DI water, and then resuspended in 25 μL of DI water. 5 μL of cell solution was mixed with nanorods in a 1:1 ratio, and 3 μL of the suspension was dried on a gold-coated glass substrate.

## RAMAN SPECTRAL DATA COLLECTION AND PROCESSING

The BCG strains were illuminated with a 633 nm laser powered at 13.17 mW. The Raman spectra were collected from 742 $cm^{-1}$ to 1802 $cm^{-1}$ using a 100x, 0.9 NA objective with a spot size of ~ 3.5 μm using the scale provided in the Horiba software, a 600 l/mm grating, and 15 second acquisition time per spectra, ensuring that spectra are collected from single to at most a few bacterial cells at a time. Python (Jupyter Notebook) was used to process spectral data. Individual spectra were smoothed using wavelet denoising, specifically using the denoise wavelet function from the scikitimage Python library. Background was then subtracted using a polynomial fit with degree 5, and spectra were individually normalized to have zero mean and unit variance across all wavenumbers.

## FABRICATION OF GOLD-COATED GLASS SUBSTRATES

The gold-coated glass substrates used in this work were prepared by evaporating a 5 nm adhesion layer of titanium, followed by 200 nm of gold at a rate of 1 A/s using the Kurt J. Lesker E-Beam Evaporator in the Stanford Nano Shared Facilities. Gold was deposited using a low deposition rate of 1 A/s to ensure high uniformity and quality and minimal roughness throughout the thin film substrate. The gold film helps minimize the background fluorescence signal from the underlying glass substrate.

## GOLD NANOROD PREPARATION FOR SERS MEASUREMENT

Hexadecyl(trimethyl)ammonium bromide (CTAB) and sodium oleate (NAOL) coated gold nanorods were synthesized following previously described protocol[37,40]. The nanorods were cleaned by centrifuging 1 mL aliquots once at (9000 rpm, 20 min) and were concentrated down to 50 µL volumes. Nanorods were mixed with bacteria in 1:1 volumetric ratio to enable uniform coating as described in our earlier work[12]. Absorption spectra were recorded using a Cary 5000 UV-vis-NIR spectrometer. Scanning electron microscopy images were taken using FEI Magellan 400 XHR Scanning Electron Microscope and the Zeiss Sigma Scanning Electron Microscope (SEM). Transmission electron microscopy images were taken using FEI Tecnai G2 F20 X-TWIN Transmission Electron Microscope (TEM). Gold nanorods with a plasmon resonance centered at 780 nm with a height of ~99 nm and a width of ~28 nm were used.

## CNN ARCHITECTURE AND CLASSIFICATION

Our CNN model is adapted from a ResNet[25] architecture, consisting of an initial convolutional layer with 64 filters, 7 residual layers, and one fully connected layer. Each residual layer consists of two residual blocks and each convolutional layer has 32 hidden nodes. We used a minibatch size of 32. To assess the classification performance, we used 10-fold cross validation, which splits the dataset into 10 equal stratified sets (Scikit-learn StratifiedKFold). For each fold i, we leave set i as the test set and train model i using all remaining data. Model i is then used to classify data in set i (test set). We repeat this process for all 10 folds, which produces unbiased classifications for all the data, and these are compared against the ground-truth to generate confusion matrices and the overall accuracy. All t-SNE projections were plotted using Scikit-learn manifold t-SNE with a perplexity = 15.

To determine the most important wavenumbers or spectral bands affecting classification performance, we used a probing algorithm that perturbs the input data before re-running the classification through the trained models. Using the test fold from our 10-fold cross validation, we iterate through the wavenumbers and at each iteration, perturb the spectrum by modulating the amplitude of the spectral intensity with a Voigt distribution centered at the probing wavenumber. After each perturbation, we recalculate the classification accuracy, compare the updated results with our baseline classification accuracy, and determine the importance for each wavenumber - the greater the decrease in accuracy due to a given perturbation, the more important the wavenumber. Each wavenumber of each spectrum in the test fold is perturbed 5 times by randomly sampling spectral intensity values at that wavenumber in order to create random, but spectrally sensible instances. Finally, all results are averaged to determine our final feature importance.

## DATA COLLECTION ON OCTOPI-RAMAN

Raman spectra were collected from 350 $cm^{-1}$ to 1606 $cm^{-1}$ using a 20x, 0.75 NA objective with spot size of 5.5 μm (Gaussian diameter; FWHM of 3.2 μm) , laser power of 2.53 mW , and 0.3 second acquisition time per spectrum. The laser spot size was determined using a stage micrometer as a standard.

## SPUTUM SAMPLES

To obtain processed sputum for Raman interrogation, sputum samples were decontaminated and concentrated per standard laboratory procedure[41] as follows: the sputum samples were treated with N-acetyl-l-cysteine–sodium hydroxide (NALC-NaOH) to break up the mucus and treated with NaOH to kill the oral flora. The pH was then neutralized with phosphate buffer and samples were concentrated with centrifugation. Only sputa with a negative acid-fast bacilli microscopy result (mycobacteria-negative) and from patients at low risk for TB were used for BCG spiking. Patient samples were de-identified prior to use in our study.

**Acknowledgements:**
The authors thank Chris Cundy, Ahmed Shuabi, Dr. Fareeha Safir, Dr. Jack Hu, Dr. Halleh Balch, and Liam Herndon for many insightful discussions. The authors thank Dr. Fareeha Safir for assistance with collecting SEM images of BCG, Liam Herndon for assistance with gold nanorod synthesis, and Ariel Stiber for assistance collecting TEM images of gold nanorods. We thank the Chan Zuckerberg Biohub San Francisco for funding, as well as the Stanford BMGF-funded Institute for Immunity, Transplantation, and Infectious Disease for seed funding. The authors would also like to thank Vida Shokoohi and Dr. John Coller at the Stanford Functional Genomics Facility for their assistance with BCG sequencing. Geneious Prime 2024.0.2 was used to analyze the BCG sequencing data. Some elements of Figure 1a were created with BioRender.com

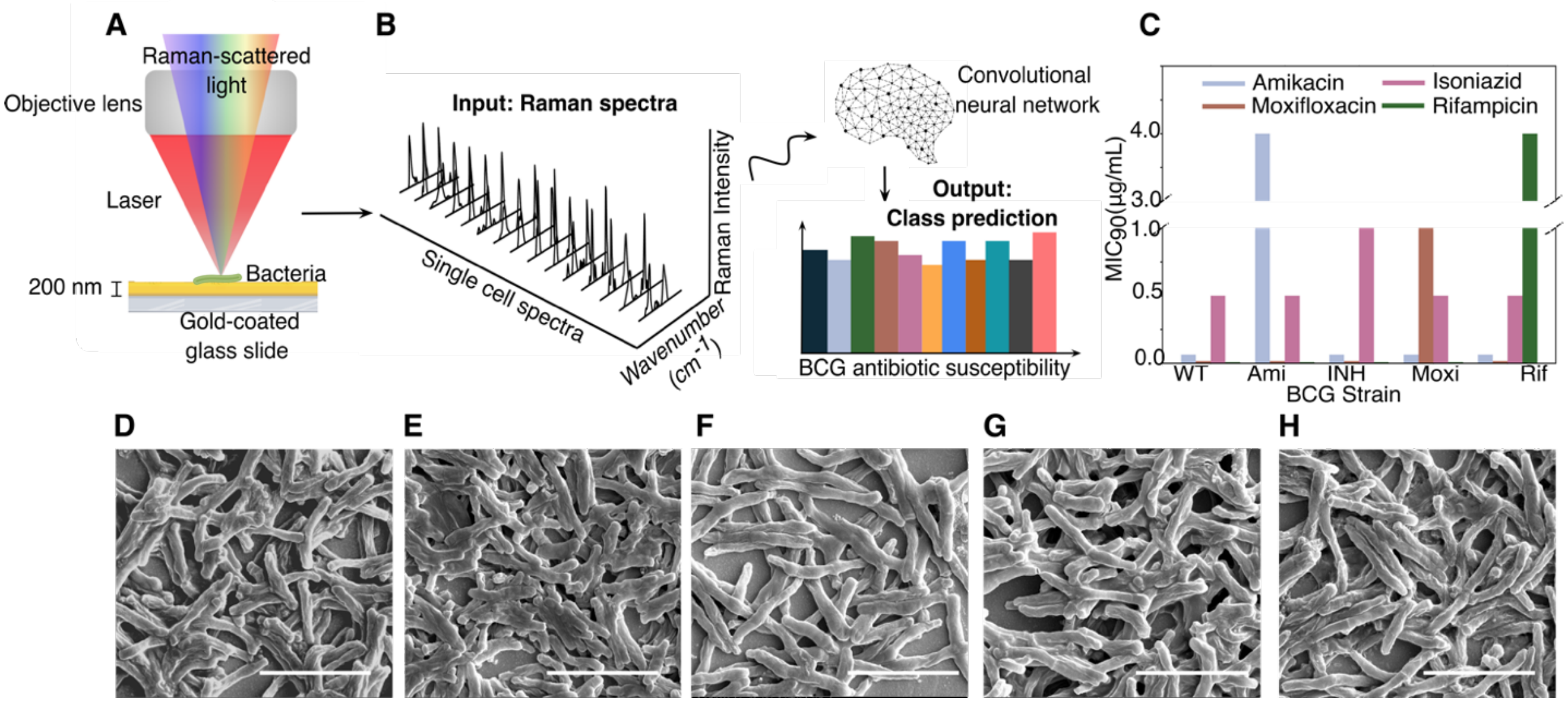

**Figure 1. Experimental setup and example bacterial Raman spectra with antibiotic resistance profiles.** A) Diffraction- limited spot of a 633 nm incident laser collecting a spectrum from roughly a single bacterial cell. B) Bacterial spectra are used as input for the neural network to perform a classification task that outputs probability scores for each bacterium antibiotic resistance. C) Minimum inhibitory concentrations for the five distinct strains tested across the four main antibiotics using standard serial two-fold dilution. See Table S1 for details. (D, E, F, G, H) Scanning electron microscopy (SEM) image of a monolayer of wildtype, amikacin-resistant, rifampicin-resistant, isoniazid-resistant, and moxifloxacin-resistant BCG. Scale bar is 4 µm.

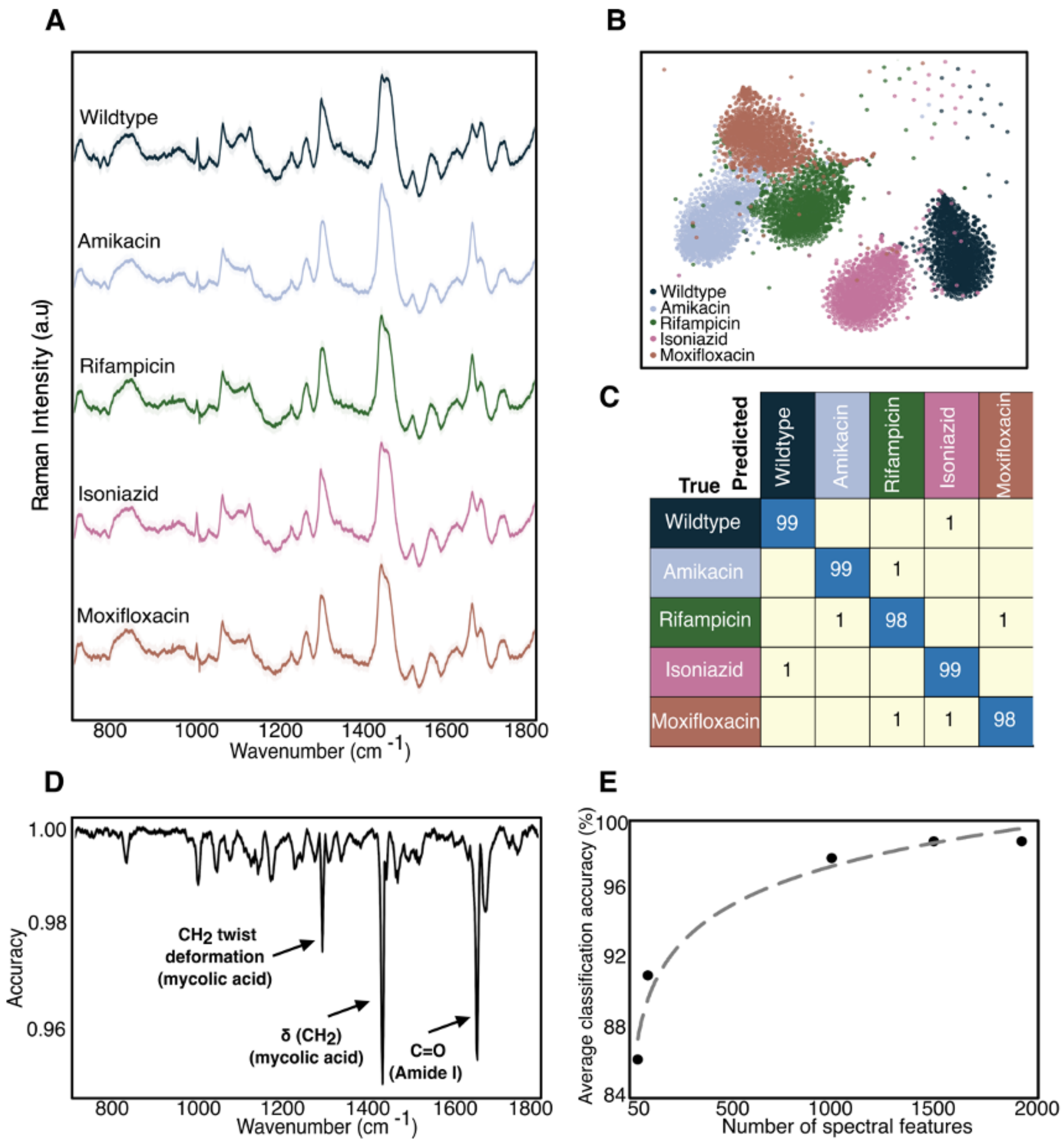

**Figure 2. BCG antibiotic resistance determination.** A) Average spectrum with standard deviation of all spectrograms (1700 spectra collected at 15s acquisition) from the 5 BCG strains. The spectra show significant differences in both peak position and peak intensity at bands centered at ~1000 $cm^{-1}$, 1060 $cm^{-1}$, and between ~1300-1700 $cm^{-1}$. B) Two-dimensional t-SNE projection across all Raman spectra of the dataset for 5 BCG samples susceptible and resistant to the four different antibiotics tested grouped according to antibiotic class showing clustering. C) Normalized confusion matrix generated using CNN on the few-to-single-cell spectra collected from 5 BCG strains, grouped by antibiotic class. Samples were evaluated by performing a stratified K-fold cross validation of our classifier's performance across 10 splits, showing ~98% classification accuracy across all samples. D) Feature selection highlighting the identification of true, physical vibrational modes. Feature selection performed to determine relative weight of spectral wavenumbers in our CNN classification. E) Average classification accuracy as a function of the number of top spectral features selected showing 96% accuracy with only 500 features.

**A**

| Sub variant strain | Gene of interest | Mutation | Position |
|---|---|---|---|
| Isoniazid 1 | katG | C → T | 609 |
| Isoniazid 2 | katG | A → G, C → T | 290, 609 |
| Isoniazid 3 | katG | G → A | 1292 |
| Isoniazid 4 | katG | None | None |

**B**

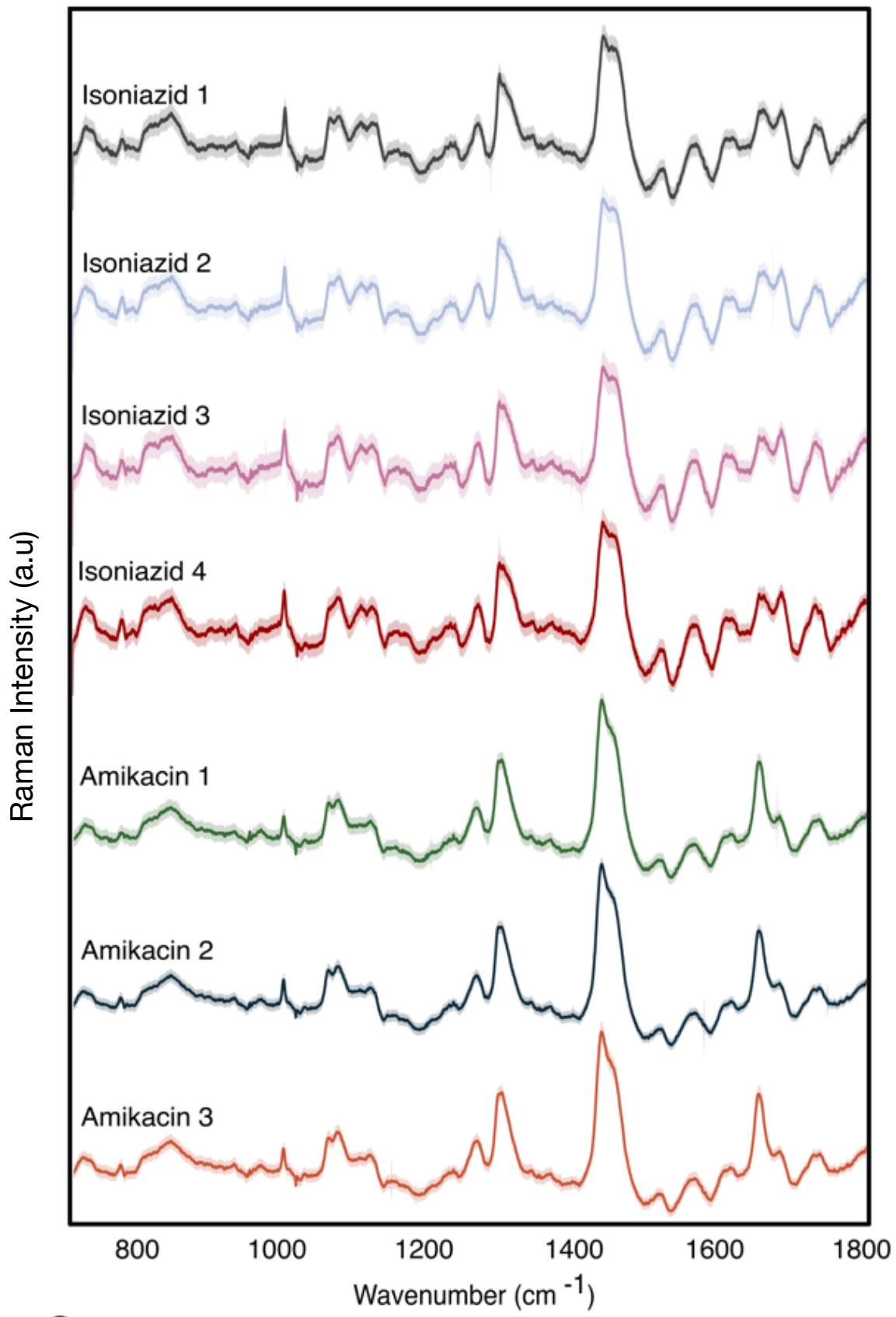

**C**

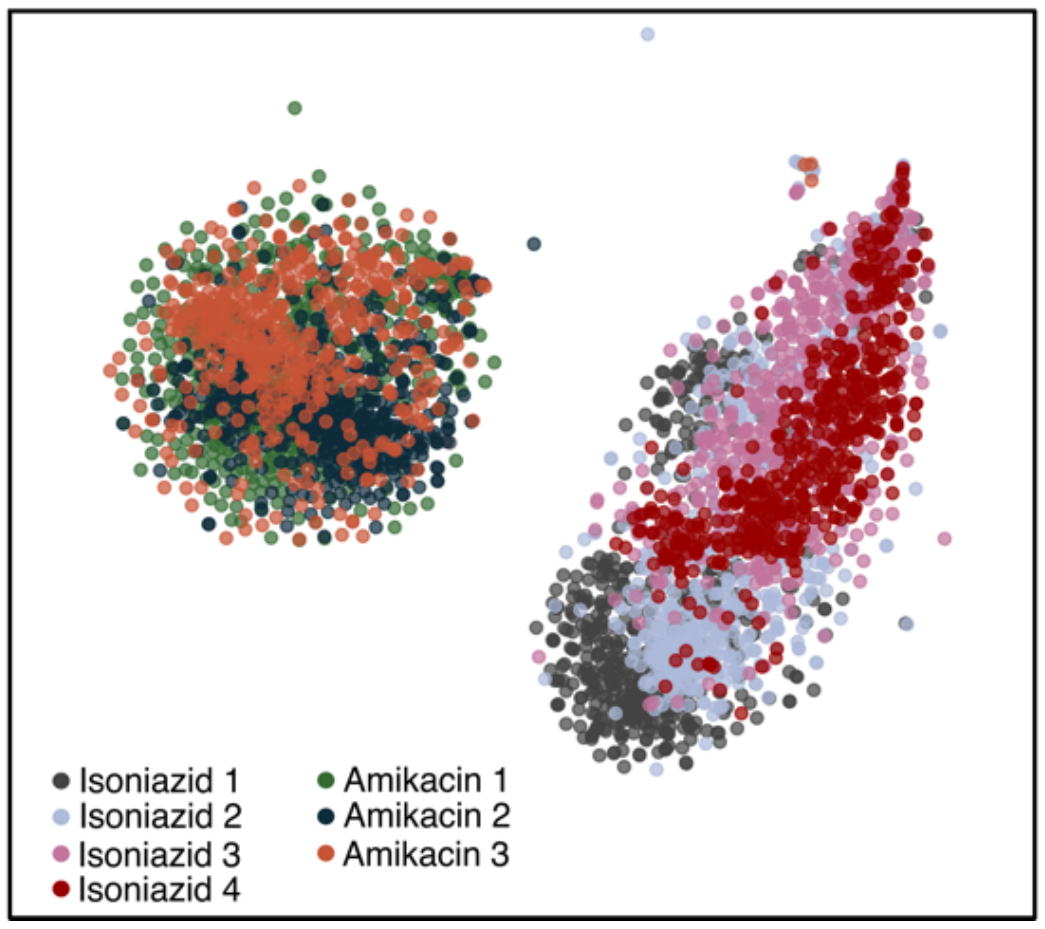

**Figure 3. Resistant mutant prediction.** A) Sequencing results showing single mutation differences in resistance genes of interest resulting in mutants of resistant strains against isoniazid and amikacin B) Average Raman spectra of 529 spectra with standard deviation of four isoniazid resistant mutants and three amikacin-resistant mutants showing nearly identical fingerprints among their respective class of resistance. C) t-SNE plot showing overlapping clusters of mutants within a class of antibiotic resistance.

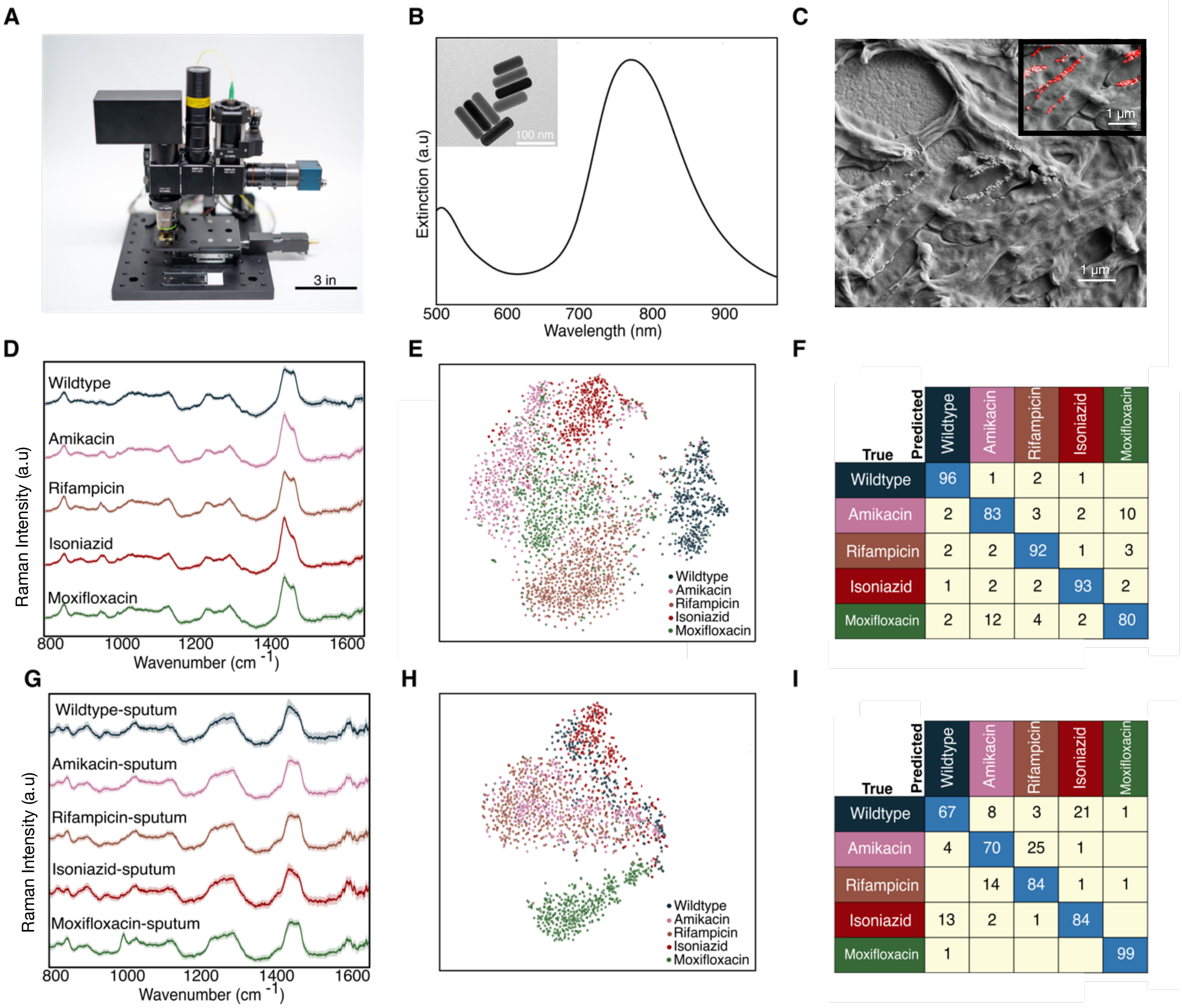

**Figure 4. Demonstration on our low-cost Raman system.** A) Photo of our system showcasing its small footprint and portable platform for use at the point-of-care. A standard microscope slide is included for scale. B) Extinction spectra of gold nanorods showing plasmon resonance centered at ~780 nm to overlap with the setup's 785 nm excitation laser. Panel (B) inset shows transmission electron micrograph of gold nanorods used. C) Scanning electron micrograph of BCG mixed with gold nanorods, which increase the Raman scattering from single cells. Panel (C) inset shows a false-colored scanning electron micrograph of gold nanorods colored red D) Raman spectra of the 5 BCG strains mixed with gold nanorods imaged on our setup. E) t-SNE projection across all Raman spectra of the 5 BCG strains. F) Average accuracy of ~89% is achieved using spectra from bacteria suspended in water and dropcast dried (replication of Fig. 2c on our setup) G) Raman spectra of the 5 BCG strains spiked in sputum imaged on our setup. H) t-SNE projection across all spiked sputum samples. I) Average accuracy of ~79% is achieved using spectra collected from two separate spiked sputum samples.

# Supporting Information for Rapid, antibiotic incubation-free determination of tuberculosis drug resistance using machine learning and Raman spectroscopy

**Authors:** Babatunde Ogunlade[1*†], Loza F. Tadesse[2,3,4*], Hongquan Li[5*], Nhat Vu[6], Niaz Banaei[7], Amy K. Barczak[4,8,9], Amr. A. E. Saleh[1,10], Manu Prakash[2] and Jennifer A. Dionne[1,11†]

**Affiliations:**
[1] Department of Materials Science and Engineering, Stanford University; Stanford, 94305, CA, USA.

[2] Department of Bioengineering, Stanford University School of Medicine and School of Engineering; Stanford, 94305, CA, USA.

[3] Department of Mechanical Engineering, Massachusetts Institute of Technology; Cambridge, 02142, MA, USA.

[4] The Ragon Institute, Massachusetts General Hospital; Cambridge, 02139, MA, USA.

[5] Department of Applied Physics, Stanford University; Stanford, 94305, CA, USA.

[6] Pumpkinseed Technologies, Inc; Palo Alto, 94306, CA, USA.

[7] Department of Pathology, Stanford University School of Medicine; Stanford, 94305, CA, USA.

[8] Division of Infectious Diseases, Massachusetts General Hospital; Boston, 02114, MA, USA.

[9] Department of Medicine, Harvard Medical School; Boston, 02115, MA, USA.

[10] Department of Engineering Mathematics and Physics, Cairo University; Giza, 12613, Egypt.

[11] Department of Radiology, Molecular Imaging Program at Stanford (MIPS), Stanford University School of Medicine; Stanford, 94035, CA, USA.

*Indicates equal contribution

† To whom correspondence should be addressed;

E-mail: bogun@stanford.edu; jdionne@stanford.edu.

**This PDF file includes:**

Figures S1 to S13
Table S1

# SI References

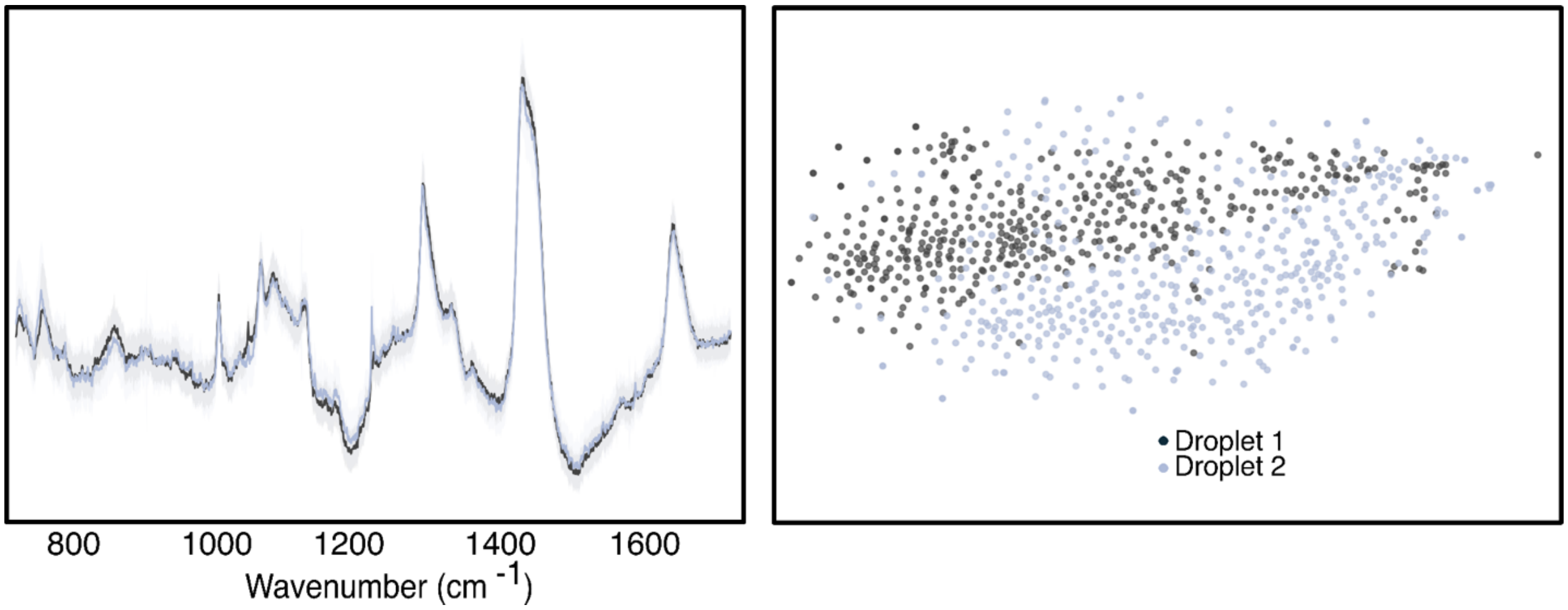

**Fig. S1.** Droplet repeatability study. 450 Raman spectra of wildtype BCG collected on two separate droplets, showing spectral overlap and T-SNE overlap.

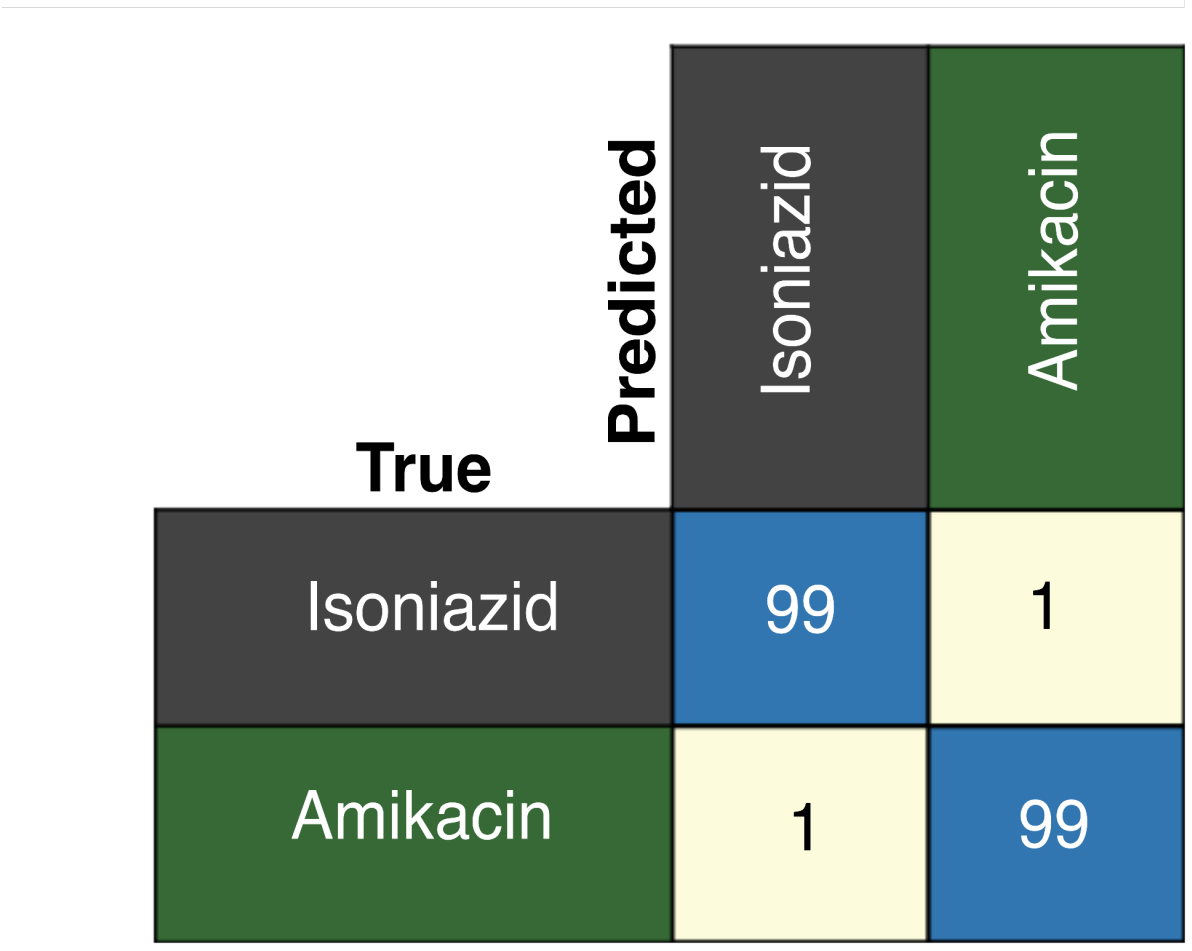

**Fig. S2.** Confusion matrix classifying the isoniazid and amikacin resistant mutants when grouped according to their antibiotic class, showing ~99% classification accuracy (529 spectra per strain; 2116 spectra for the Isoniazid grouping and 1587 spectra for the Amikacin grouping)

| True \ Predicted | Wildtype | Amikacin | Rifampicin | Isoniazid | Moxifloxacin |
|---|---|---|---|---|---|
| Wildtype | 95 | | | 5 | |
| Amikacin | | 97 | 1 | | 2 |
| Rifampicin | 1 | 1 | 95 | 1 | 2 |
| Isoniazid | 5 | | | 94 | 1 |
| Moxifloxacin | | 3 | 2 | | 95 |

| True \ Predicted | Wildtype | Amikacin | Rifampicin | Isoniazid | Moxifloxacin |
|---|---|---|---|---|---|
| Wildtype | 96 | | | 4 | |
| Amikacin | | 98 | | | 2 |
| Rifampicin | 1 | 1 | 95 | 1 | 2 |
| Isoniazid | 5 | | | 95 | |
| Moxifloxacin | | 4 | 2 | 1 | 93 |

| | Wildtype | Amikacin | Rifampicin | Isoniazid | Moxifloxacin |
|---|---|---|---|---|---|
| Wildtype | 98 | | | 1 | 1 |
| Amikacin | | 97 | | 1 | 2 |
| Rifampicin | | | 99 | 1 | |
| Isoniazid | | | 1 | 97 | 2 |
| Moxifloxacin | | 2 | | 4 | 94 |

| | Wildtype | Amikacin | Rifampicin | Isoniazid | Moxifloxacin |
|---|---|---|---|---|---|
| Wildtype | 96 | | 1 | 1 | 2 |
| Amikacin | | 93 | 1 | 5 | 1 |
| Rifampicin | 1 | | 96 | 2 | 1 |
| Isoniazid | 1 | 9 | 2 | 82 | 6 |
| Moxifloxacin | 3 | 3 | 1 | 7 | 86 |

**Fig. S3.** Confusion matrices from four replicate studies on the Horiba Labram scientific grade Raman microscope showing an average of ~95% classification accuracy across the 5 BCG strains ( ~450, 529, ~420, and ~420 spectra per strain collected, respectively).

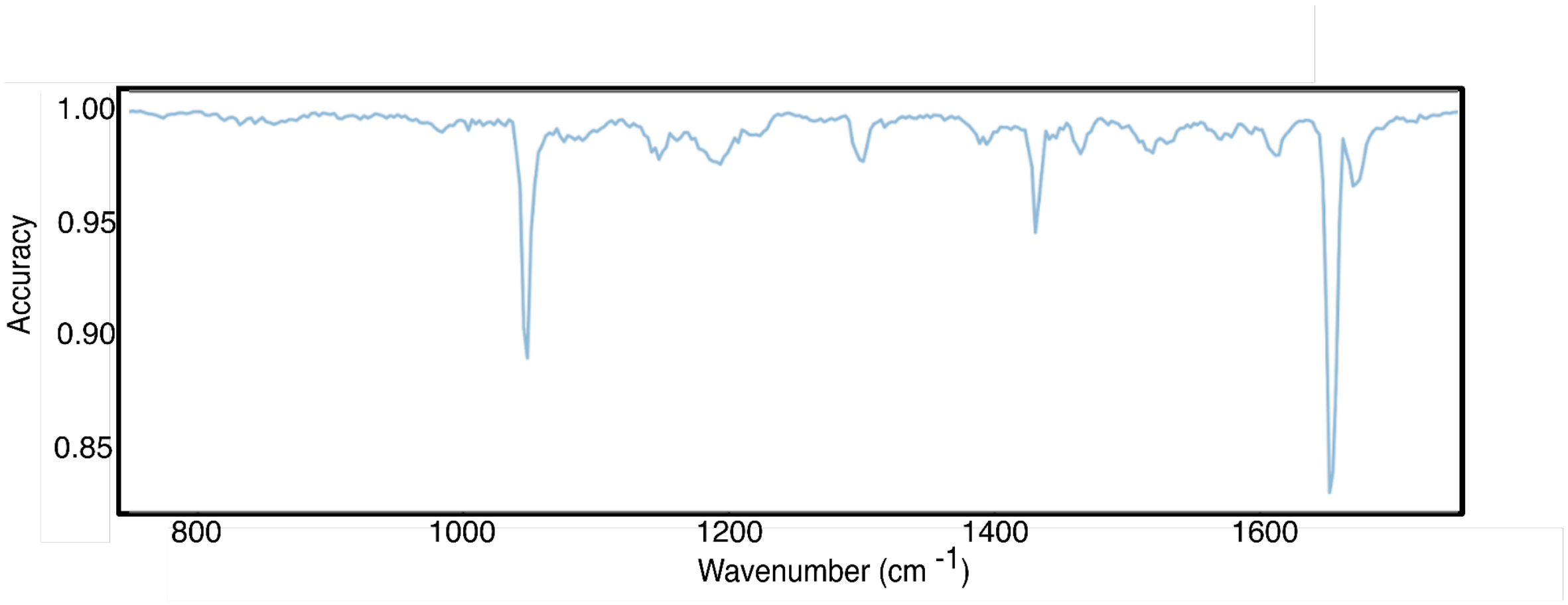
1.00
0.95
0.90
0.85
Accuracy
800
1000
1200
1400
1600
Wavenumber (cm -1)

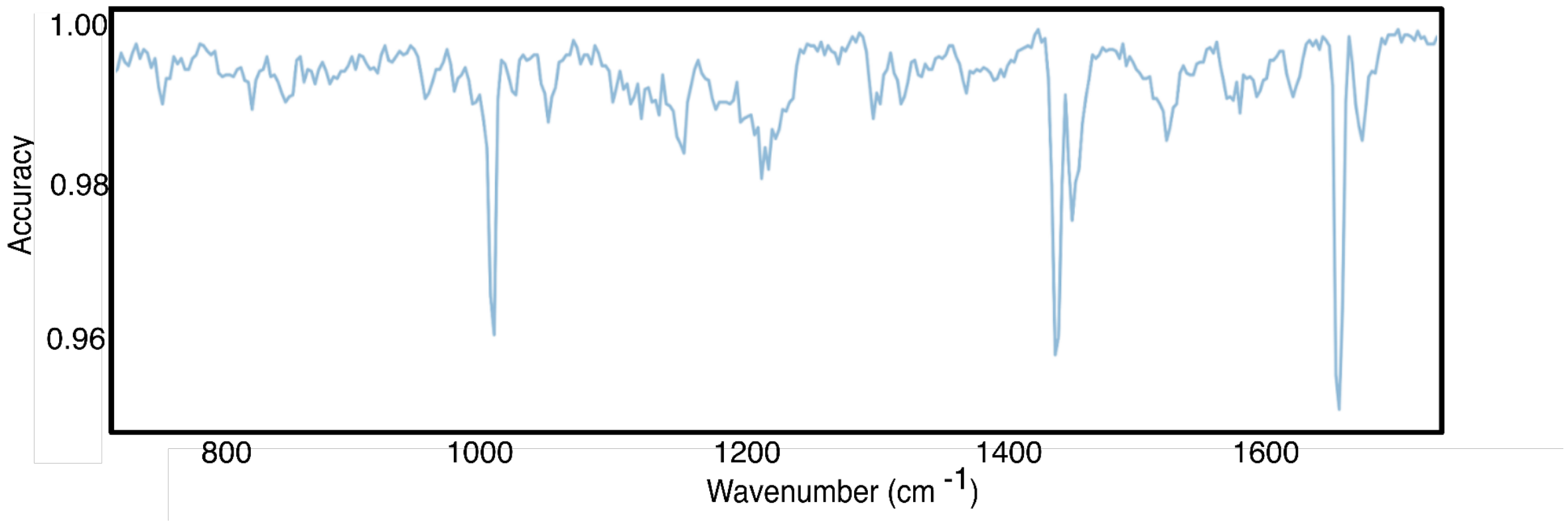
1.00
0.98
0.96
Accuracy
800
1000
1200
1400
1600
Wavenumber (cm -1)

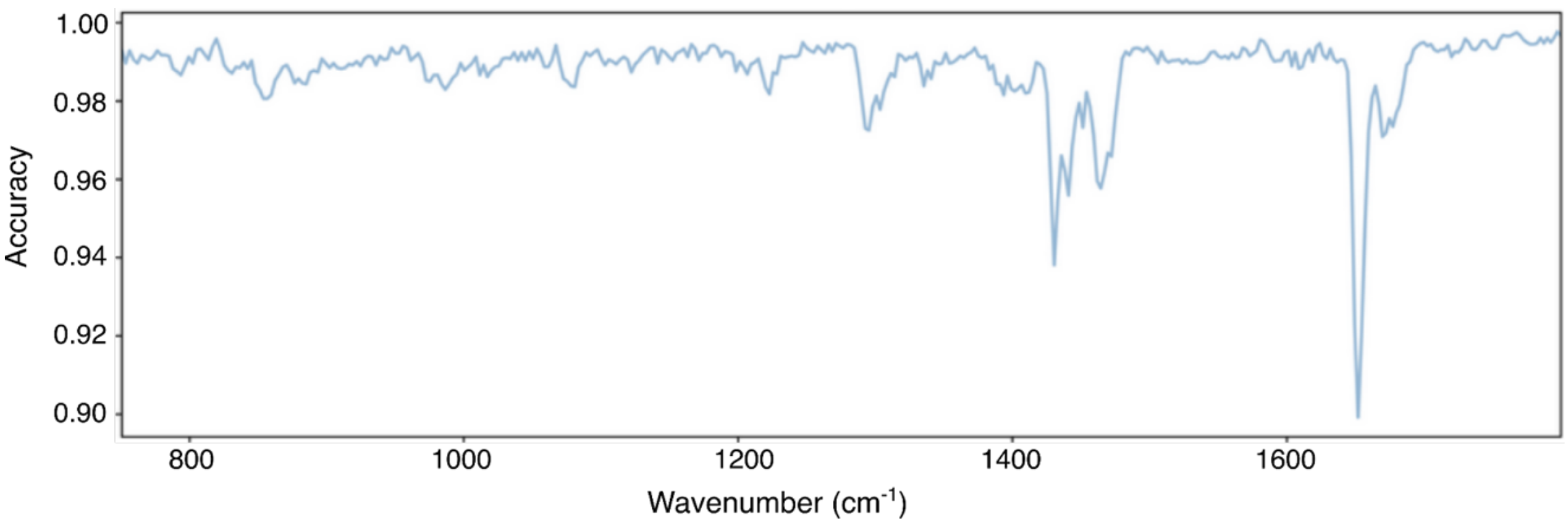
1.00
0.98
0.96
0.94
0.92
0.90
Accuracy
800
1000
1200
1400
1600
Wavenumber (cm-1)

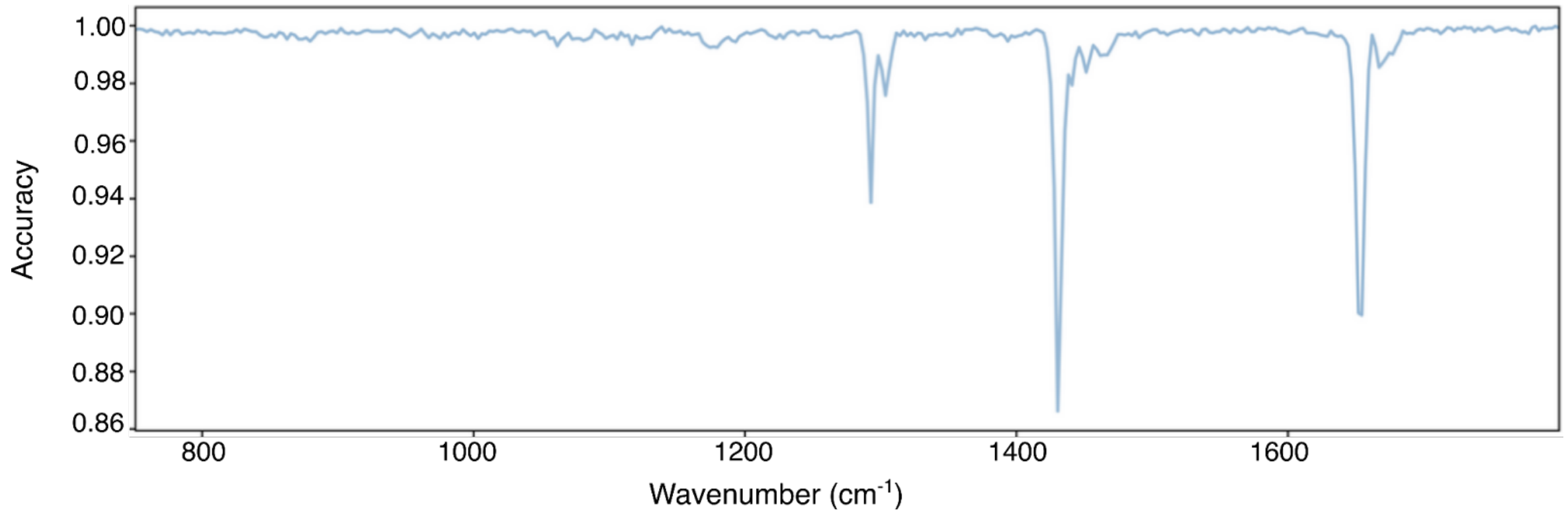

**Fig. S4.** Feature selection from four replicate studies on the Labram.

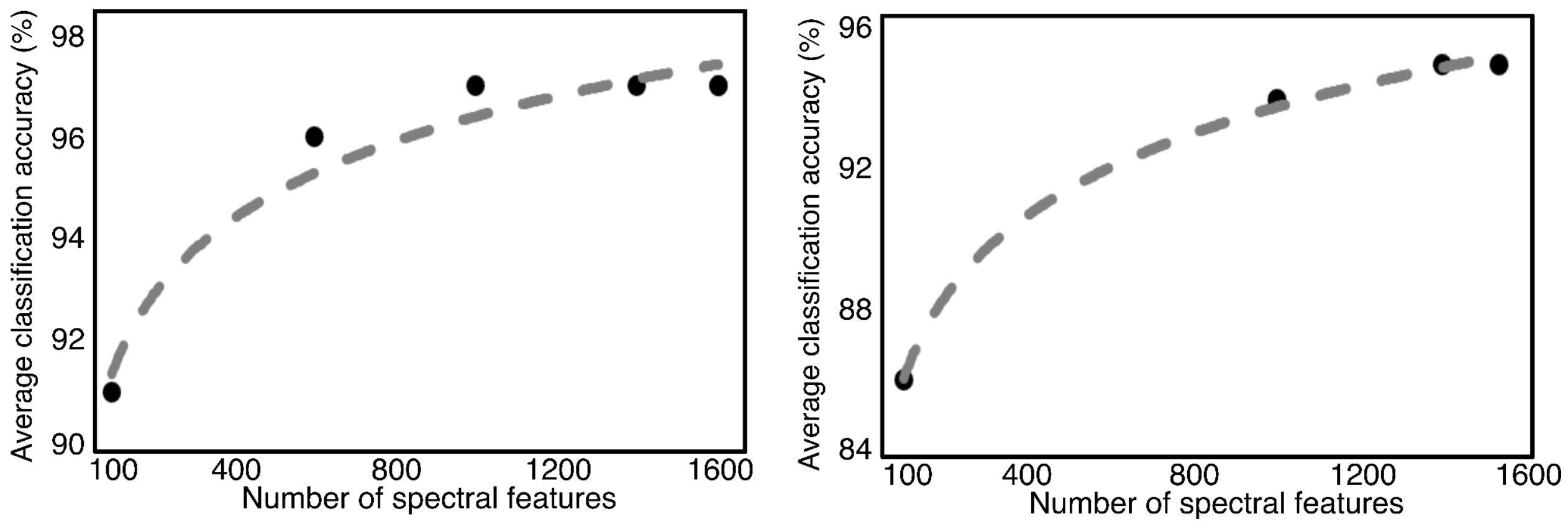

**Fig. S5.** Average classification accuracy vs number of spectral features for two replicate Labram studies.

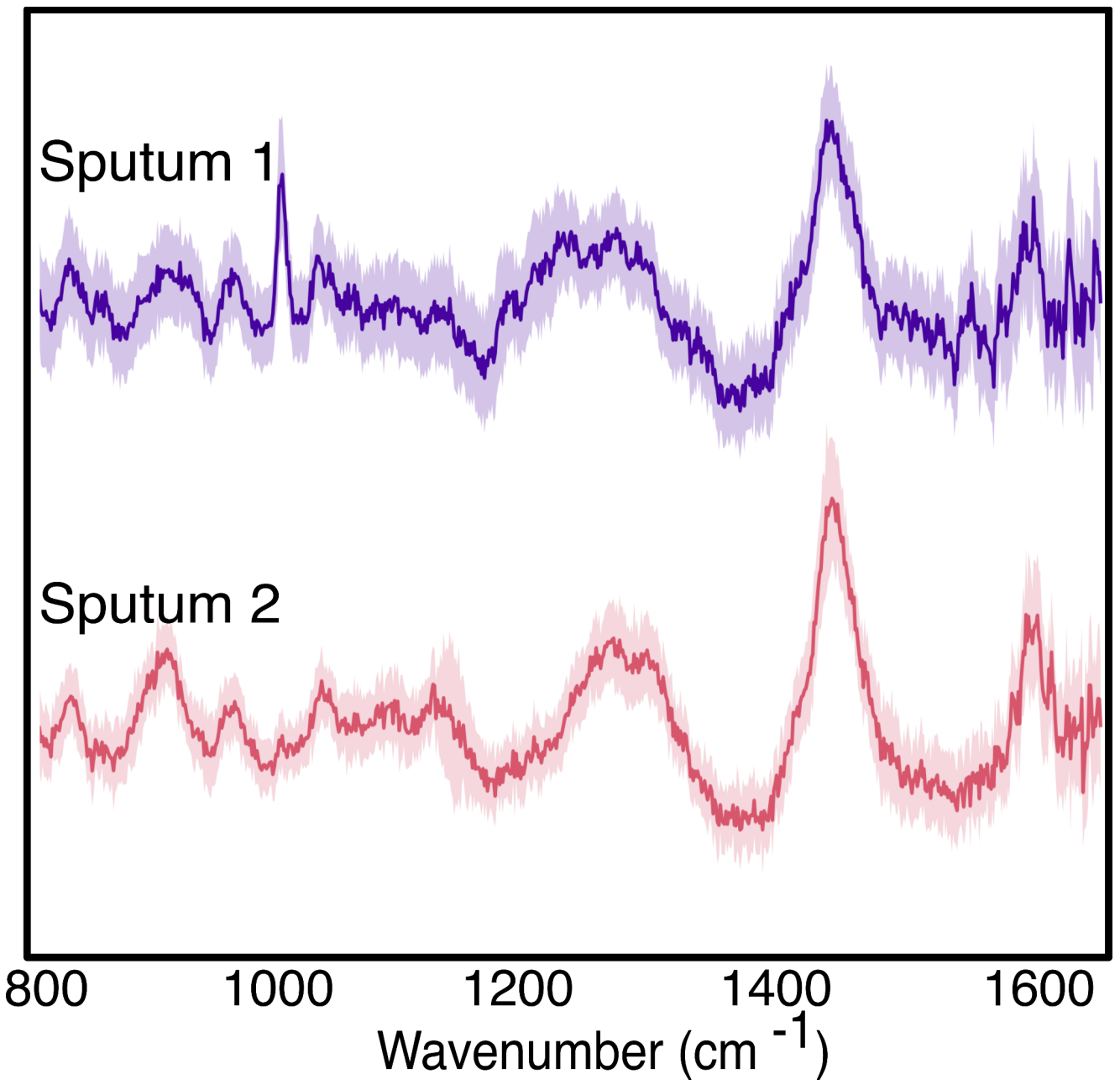

**Fig. S6.** Average Raman spectra of the two sputum samples used in the Octopi study.

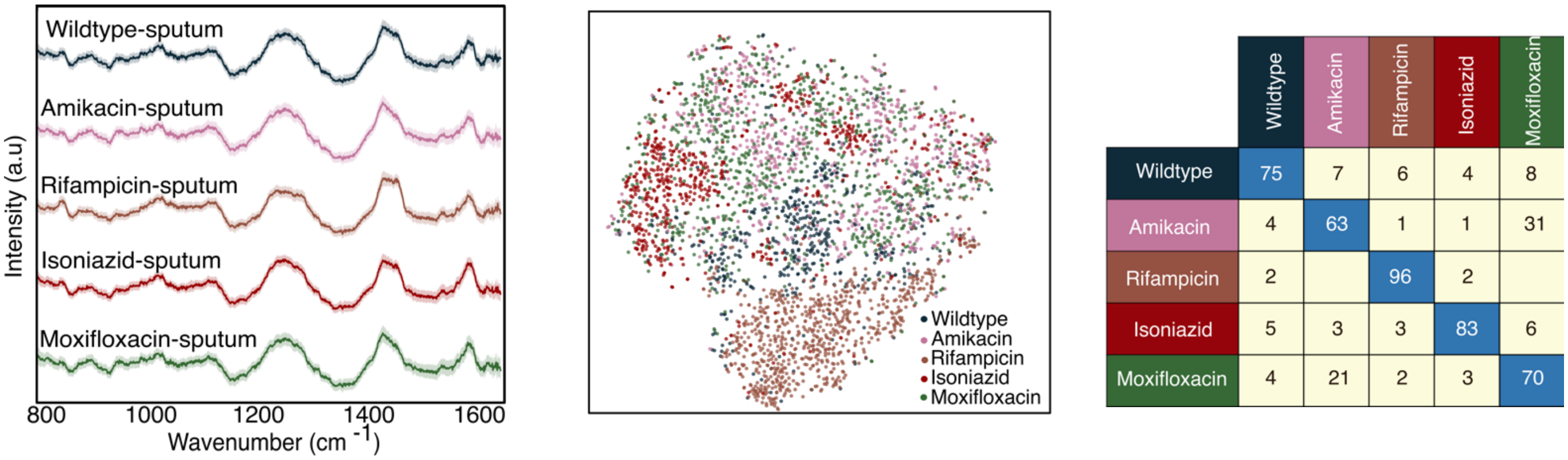

| | Wildtype | Amikacin | Rifampicin | Isoniazid | Moxifloxacin |
|---|---|---|---|---|---|
| Wildtype | 75 | 7 | 6 | 4 | 8 |
| Amikacin | 4 | 63 | 1 | 1 | 31 |
| Rifampicin | 2 | | 96 | 2 | |
| Isoniazid | 5 | 3 | 3 | 83 | 6 |
| Moxifloxacin | 4 | 21 | 2 | 3 | 70 |

**Fig. S7**. Mean Raman spectra of 5 BCG strains spiked in other sputum sample collected on Octopi-Raman with t-SNE showing some clustering by antibiotic class, and the confusion matrix demonstrating ~ 77% average classification accuracy.

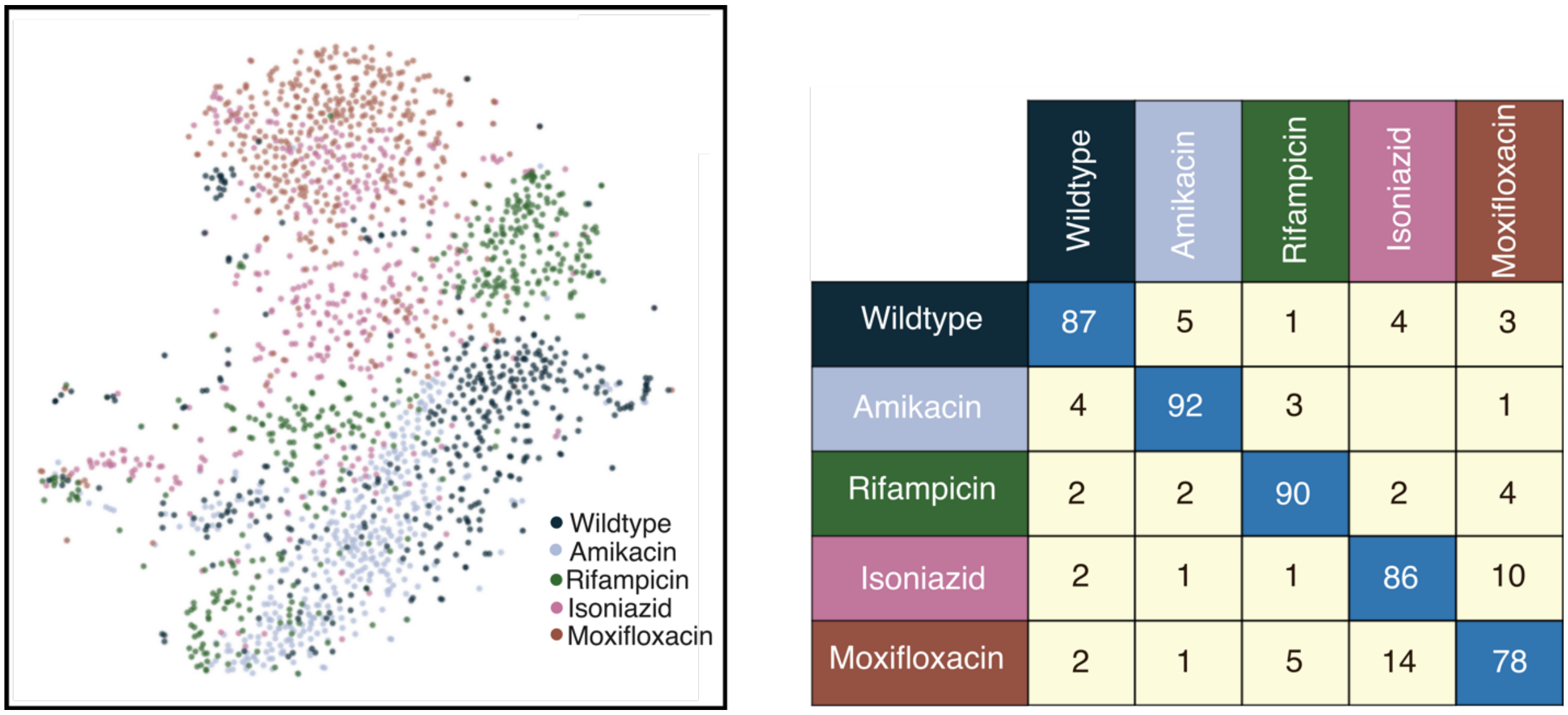

| | Wildtype | Amikacin | Rifampicin | Isoniazid | Moxifloxacin |
|---|---|---|---|---|---|
| Wildtype | 87 | 5 | 1 | 4 | 3 |
| Amikacin | 4 | 92 | 3 | | 1 |
| Rifampicin | 2 | 2 | 90 | 2 | 4 |
| Isoniazid | 2 | 1 | 1 | 86 | 10 |
| Moxifloxacin | 2 | 1 | 5 | 14 | 78 |

**Fig. S8** T-SNE of 5 BCG strains spiked in Sputum 2 collected on the Labram showing clear clustering by antibiotic class, and the confusion matrix demonstrating ~87% average classification accuracy.

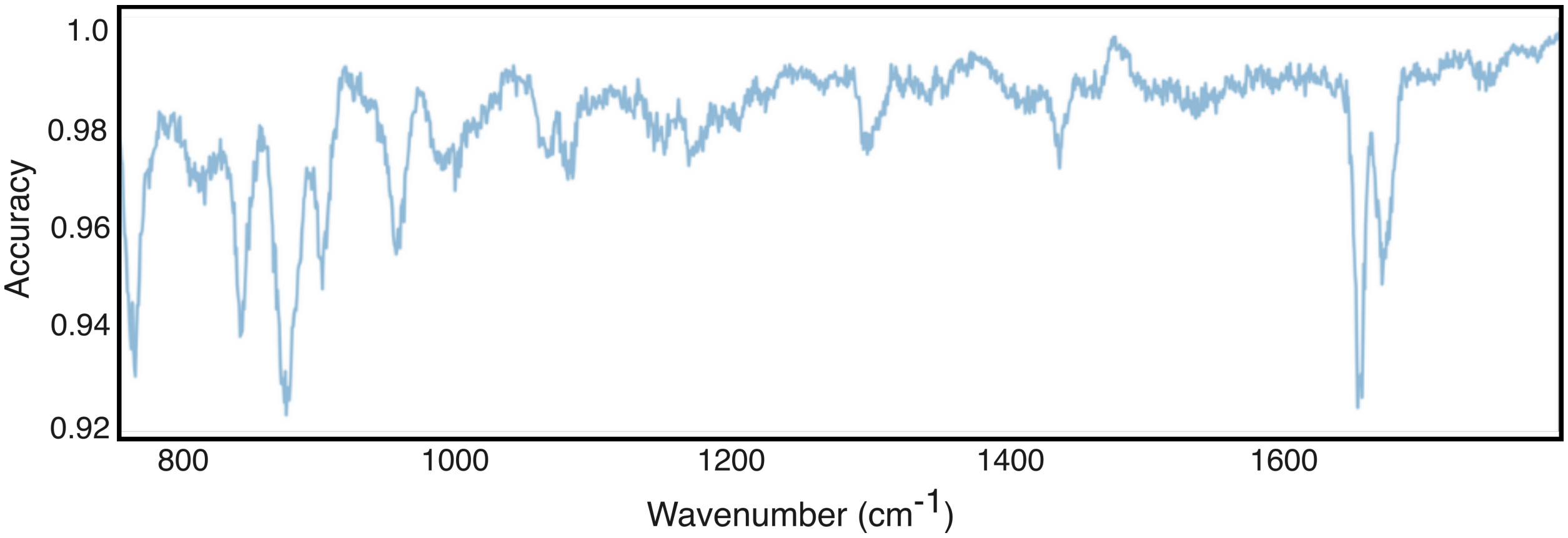

**Fig. S9** Feature selection of 5 BCG strains spiked in Sputum 2 collected on the Labram

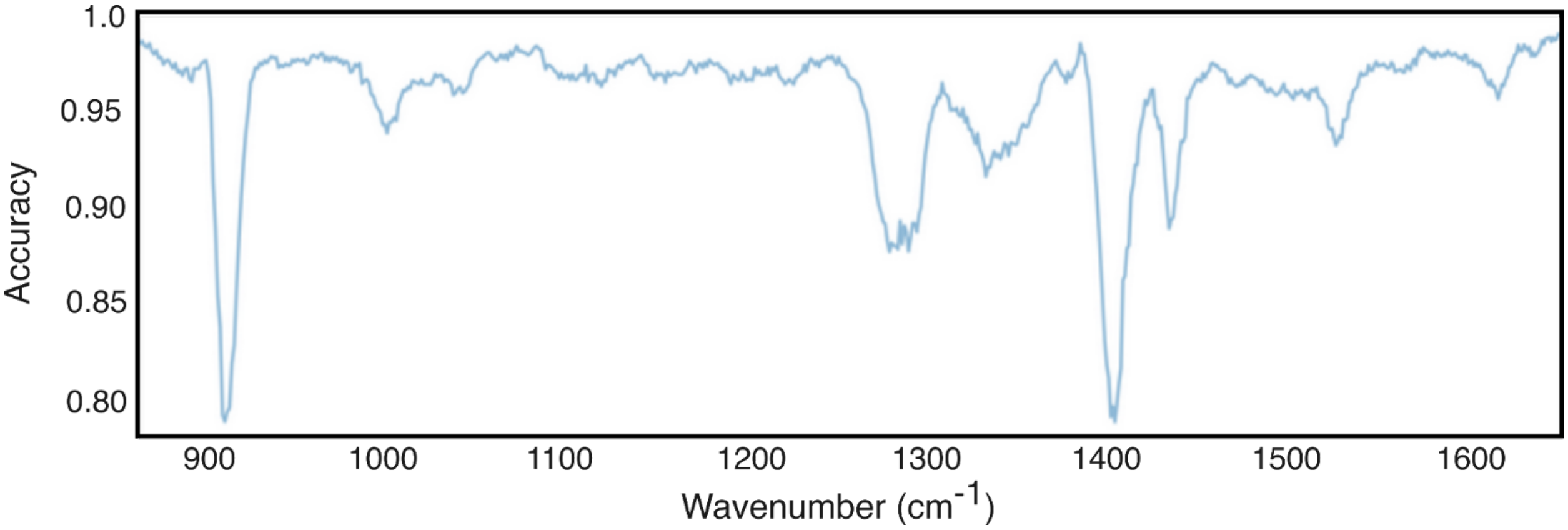

**Fig. S10** Feature selection of 5 BCG strains collected on Octopi-Raman

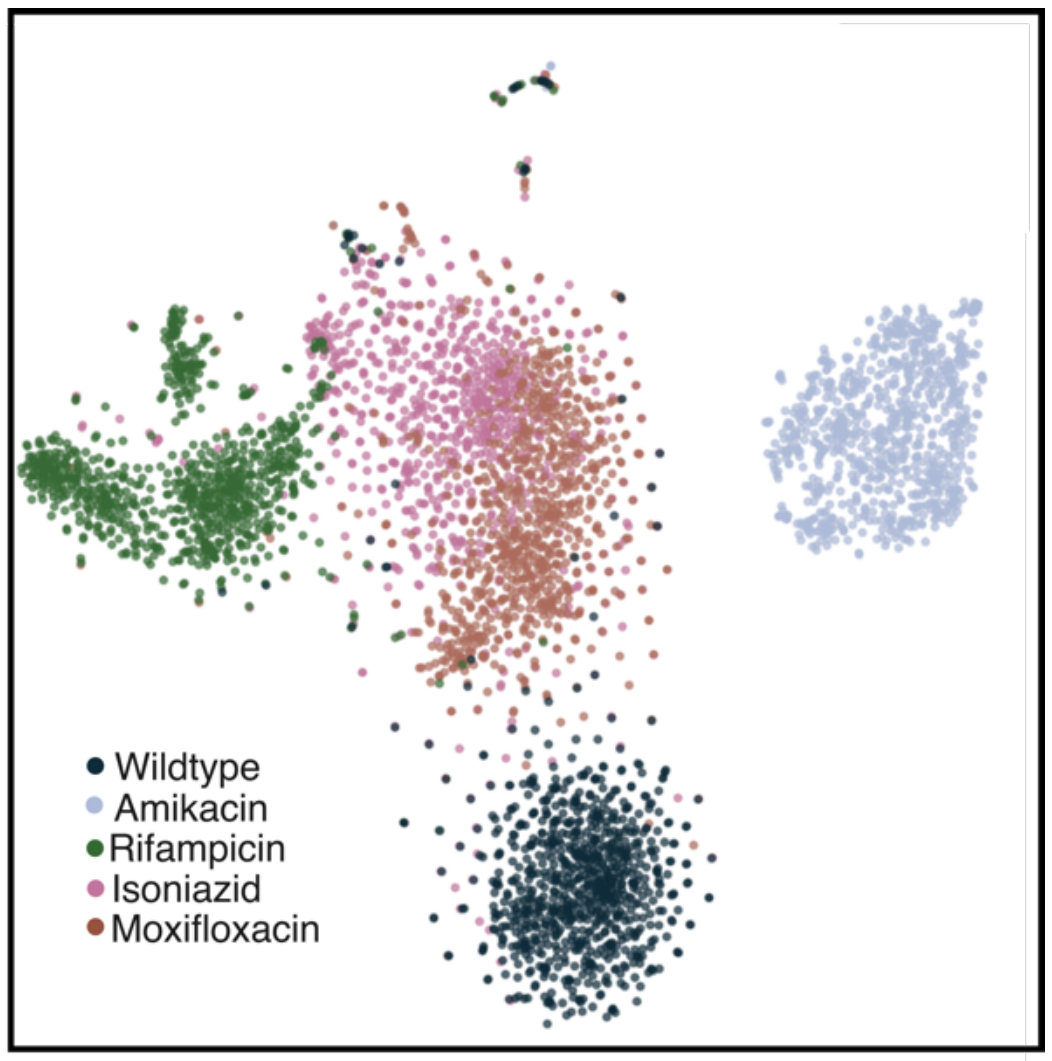

| | Wildtype | Amikacin | Rifampicin | Isoniazid | Moxifloxacin |
|---|---|---|---|---|---|
| Wildtype | 97 | | | 1 | 2 |
| Amikacin | | 99 | | 1 | |
| Rifampicin | 1 | | 95 | 2 | 2 |
| Isoniazid | 1 | | 2 | 87 | 10 |
| Moxifloxacin | 1 | | 2 | 10 | 87 |

**Fig. S11** T-SNE and confusion matrix (~840 spectra/ class) of combined datasets from 5 BCG strains from two experimental replicates

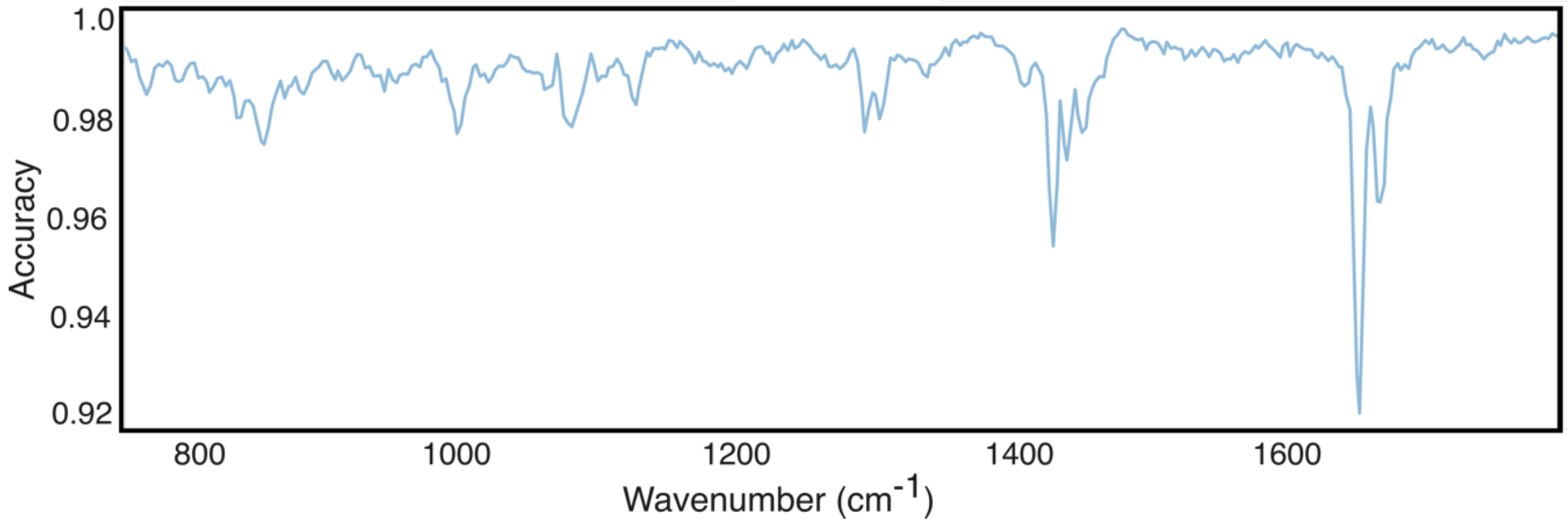

**Fig. S12** Feature selection of combined datasets from 5 BCG strains from two experimental replicates

| BCG strain | Gene of interest | Mutation | Position |
|---|---|---|---|
| Isoniazid 1 | katG | C-->T | 609 |
| Isoniazid 2 | katG | A-->G, C-->T | 290, 609 |
| Isoniazid 3 | katG | G-->A | 1292 |
| Isoniazid 4 | katG | none | none |
| Moxifloxacin 1 | GyrB | C-->T, C-->T | 1167, 1207 |
| Amikacin 1 | rrs | A-->G | 1401 |
| Amikacin 2 | rrs | A-->G | 1401 |
| Amikacin 3 | rrs | A-->G | 1401 |
| Rifampicin | rpoB | T-->C, C-->T | 1325, 1349 |

**Fig S13** List of SNPs present in antibiotic-resistance associated genes of interest for each strain[42]

**Table S1:**

| MIC$_{90}$ ($\mu g/mL$) | | | | | |
|---|---|---|---|---|---|
| Antibiotic | Wildtype BCG | Amikacin-resistant BCG | Isoniazid-resistant BCG | Moxifloxacin-resistant BCG | Rifampicin-resistant BCG |
| Amikacin | 0.0625 | $>$ 4.0 | 0.0625 | 0.0625 | 0.0625 |
| Isoniazid | 0.5 | 0.5 | $<$1 | 0.5 | 0.5 |
| Moxifloxacin | 0.016 | 0.016 | 0.016 | 1 | 0.016 |
| Rifampicin | $<$0.008 | $<$0.008 | $<$0.008 | $<$0.008 | $<$4.0 |

**Table S1:** Minimum inhibitory concentrations of the five main BCG strains studied. This was obtained using standard serial two-fold dilution method.

**References (40, 41, 42):**

**Funding:**

- Stanford Catalyst for Collaborative Solutions (funding ID 132114)
- Stanford BMGF-funded Institute for Immunity, Transplantation, and Infectious Disease (grant number OPP 1113682)
- The Chan Zuckerberg Biohub San Francisco Investigator Program
- The Gates Foundation (grant number OPP 1113682)
- The National Science Foundation (grant number 1905209)
- The NIH New Innovator Award (1DP2AI152072-01).
- Stanford Graduate Fellowship
- Stanford EDGE Fellowship
- Stanford Bio-X Bowes Graduate Student Fellowship
- Part of this work was performed at the Stanford University Nano Shared Facilities and Stanford University Nanofabrication Facilities, which are supported by the National Science Foundation and National Nanotechnology Coordinated Infrastructure under awards ECCS-2026822 and ECCS-1542152 as well as the Stanford University Cell Sciences Imaging Core Facility.

**Data and materials availability:**

All data needed to evaluate the conclusions in the paper are present in the paper and/or the Supplementary Materials.